# Prognostic watch of the electric power system

S. Z. Stefanov

*ESO EAD, 8 Triaditsa Str., 1040 Sofia, Bulgaria*

e-mail: szstefanov@ndc.bg

Abstract. A prognostic watch of the electric power system (EPS) is framed up, which detects the threat to EPS for a day ahead according to the characteristic times for a day ahead and according to the droop for a day ahead. Therefore, a prognostic analysis of the EPS development for a day ahead is carried out. Also the power grid, the electricity market state, the grid state and the level of threat for the power grid are found for a day ahead. The accuracy of the built up prognostic watch is evaluated.

Keywords: electric power system; threat; watch; prognostics; power grid; electricity market

## 1. Introduction

The market watch (Sobajic & Douglas, 2004) is an observer of the electricity market characteristics, of the grid states and of the electricity market states. The market watch is introduced (Sobajic & Douglas, 2004) in order to integrate the market and the grid and in this way to ensure an adaptive reliability of the EPS.

The power grid states are: normal, restorative, emergency (Sobajic & Douglas, 2004). The same are the electricity market states (Sobajic & Douglas, 2004). The grid state as well the market state show threat against EPS (Kirschen & Bouffard, 2009), (Ghosh, Sharman, Rao, & Upadhyaya, 2007).

The purpose of this paper is an EPS prognostic watch construction. Herein the prognostic watch is an EPS market watch that finds out the threat against EPS for a day ahead. This prognostic watch carries out the prognostics (Schwabacher & Goebel, 2007) of the threat against EPS according to EPS characteristic times for a day ahead and according to the EPS droop for a day ahead.

Determination of the threats against EPS makes EPS more secure and more efficient. Therefore constructing the prognostic watch is a part of the project about SmartGrid (National Electrical Manufacturers Association [NEMA], 2009).

## 2. Lyapunov exponents for a day ahead

EPS development for a day ahead is characterized by the half-time replica of the morning load for day ahead $T_{6,1}$, the half-time synchronization afternoon load for a day ahead $T_{6,2}$, the time load for a day ahead $T_{24}$, the time for exchanges for a day ahead $T_{16}$ and the droop $k_c$ for a day ahead. Herein is assumed that the expected tomorrow's EPS behaviour repeats today's morning behaviour and that the expected tomorrow afternoon's EPS behaviour is synchronized with today's afternoon behaviour. The half time of the reply of the morning load, for a day ahead $T_{6,1}$, is approximately 6 hours, as the reply is not perfect. Synchronization half-time of the afternoon load for day ahead $T_{6,2}$ is approximately 6 hours because the synchronization is not perfect. EPS loading time for a day ahead $T_{24}$ is approximately 24 hours, but not just 24 hours, because it includes the expected deviation from the synchronous time. The time for exchanges $T_{16}$ is approximately 16 hours, but not just 16 hours, because of the uncertainty of the exchanges.

Let the EPS development time is a permanental field from Le Jan, Marcus, and Rosen (2012). Then the time density is a density of the potential of the transient Markov process. Let the EPS development time permanental field is determined by the number of the 4-paths among the times ($T_{6,1}$, $T_{6,2}$, $T_{16}$, $T_{24}$). Herein the number of the 4-paths is according to Benjamin and Cameron (2005). Then the time density is estimated by the permanent per(A) of the following matrix

$$A = \begin{vmatrix} T^*_{6,1} & T^*_{6,2} & 1 & 0 \\ T^*_{24} & T^*_{16} & T^*_{6,2} & 1 \\ T^*_{16} & T^*_{24} & T^*_{16} & T^*_{6,2} \\ T^*_{6,2} & T^*_{16} & T^*_{24} & T^*_{6,2} \end{vmatrix} \qquad (1)$$

$T^*_{6,1} = \{\ 2T_{6,1}/10,\ \text{if}\ T_{6,1} < 9.5;\ T_{6,1}/10,\ \text{if}\ T_{6,1} \geq 9.5\ \},\ T^*_{6,2} = T_{6,2}/10$

$T^*_{16} = \{\ 2T_{16}/10,\ \text{if}\ T_{16} < 9.5;\ T_{16}/10,\ \text{if}\ T_{16} \geq 9.5\ \},\ T^*_{24} = T_{24}/10$

This time density of the EPS development determines the following Lyapunov exponent for a day ahead

$$L_p = \ln^2(\text{per}(A))/10 + 1 \qquad (2)$$

The droop characterizes the EPS load change, when the EPS frequency is changing. Let the droop for a day ahead is assessed by the hourly volume of the exchanges. Then the droop for a day ahead may be considered as a random variable with a free Poisson distribution. The Lyapunov exponent for a random variable with a free Poisson distribution is obtained in Kargin (2008). That's why the EPS droop determines the following Lyapunov exponent for a day ahead

$$L_y = \exp(k_c/10) + 1 \qquad (3)$$

**3. Power grid for a day ahead**

The replica of today's EPS morning load creates the tomorrow's morning load. This creation is Lyapunov exponent $L_p$. Synchronization of today's afternoon EPS load annihilates the diversity of the tomorrow's afternoon loads. This annihilation is with a Lyapunov exponent $L_y$. So the replica and synchronization of the today's load create neighborhood of hourly tomorrow loads.

In such an EPS load for a day ahead, the power grid is with energy $E_1$ and angular frequency $\omega_1$, and time $t_1$, defined by Schott and Staples (2011),

$$E_1 = L_p^2, \omega_1 = 2L_p/t_1, t_1 = (1/4)(1+((L_y + L_p)/2)((L_y + L_p)/2)) \qquad (4)$$

The uncertainty of the exchanges for a day ahead and the deviation from the synchronous time for a day ahead are almost minimal for EPS with a linear stability factor $\rho$,

$$\rho = ((2+L_p)+((2+L_p)^2 - 4((4-(2+L_p)^2)/2 - 2)^{1/2})/2 \qquad (5)$$

Herein EPS is considered as Hamiltonian system. Exchanges uncertainty for a day ahead and the deviation from the synchronous time for a day ahead, are almost minimal in sense that the chaotic EPS behaviour is almost completely reduced. Then, the EPS stability domain is almost maximal around the nominal circular orbit. The linear stability factor from (5) leads to this almost maximal stability domain (Wan & Cary, 2001).

The power grid, with such an EPS stability for a day ahead, has energy $E_2$ and angular frequency $\omega_2$, and time $t_2$,

$$E_2 = (\rho + (\rho^2 - 4)^{1/2})/2, \omega_2 = 2L_y/t_2, t_2 = 5(\rho - (\rho^2 - 4)^{1/2}) \qquad (6)$$

**4. Power grid prognostic analysis for a day ahead**

The power grid is with two times that are the time $t_1$ from (4) and time $t_2$ from (6). These times are the 'frequency time' of the power grid and 'energy time' of the power grid. Then (Foster & Müller, 2010) the energy distribution in the power grid is determined by ultra-hyperbolic operator. Energy distribution in this grid is characterized by the 'frequency' solution and from 'energy' solution of the ultra-hyperbolic partial differential solution with impulse initial conditions.

The following potential $U_s$ is an energy solution of the ultra-hyperbolic equation

$$U_s = L^2_y v_1 + w_1, \quad w_1 = (3/L_p)\ln(((L_p+t_1)^2+(1/4)^2)/((L_p-t_1)^2+(1/4)^2)), \quad (7)$$

$$v_1 = (L_p+t_1)/(L_p((L_p+t_1)^2+(1/4)^2)) + (L_p-t_1)/(L_p((L_p-t_1)^2+(1/4)^2))$$

This potential is a solution of the ultra-hyperbolic equation (Kostomarov, 2006), when the probability of closeness of the EPS development to the development determined by $L_p$, is with a limit value ¼.

The following potential $U_p$ is the frequency solution of the ultra-hyperbolic equation

$$U_p = -(1/2+1/(4v_1))(1+p_x v_1/t_1)\exp(v_1 t_1), \quad p_x = 2E_1 - (\omega^2_1 + \omega^2_2) - 4 \quad (8)$$

This potential is a solution of the ultra-hyperbolic equation (Kostomarov, 2006) when the value $(E_1)^{½}$ is considered (Fulling, 2002) as a frequency.

### *4.1. Market state for a day ahead*

Energy distribution of the power grid for a day ahead sets the distance between the grid behaviour for a day ahead and its critical behaviour.

The distance between these two behaviours for the elliptical M. Riesz kernel is

$$R_e = \Gamma(1/2)/(\pi^{½} 2^6 \Gamma(3)(U_s - U_p)^{½}) \quad (9)$$

The distance between these two behaviours for M. Riesz' ultra-hyperbolic kernel is

$$R_h = 2(\omega^2_1 + \omega^2_2 + E^2_1 - E^2_2 - t^2_1)^{½}/(32\pi^2 \Gamma(3)\Gamma(3/2)) \quad (10)$$

These distances are obtained from the results (Trione, 2000) for M. Riesz' elliptic and ultra-hyperbolic kernels. Herein $\Gamma(.)$ is the gamma function.

Critical behavior of the power grid sets the following critical distance

$$R_c = \exp(-v_1 L_p)/(10 L_p) \qquad (11)$$

This distance is determined by the results (Kostomarov, 2006) for ultra-hyperbolic equation with an impulse $v_1$ from (7) in the initial moment.

The market state for a day ahead $s_m$ is

$$s_m = \begin{cases} \text{normal (n), if } R_e < R_c \text{ and } R_h < R_c \\ \text{restorative (r), if } R_e > R_c \text{ or } R_h \\ \text{emergency (e), if } R_e > R_c \text{ and } R_h > R_c \end{cases} \qquad (12)$$

*4.2. Power grid state for a day ahead*

The grid behaviour for a day ahead is critical if an edge of it is deleted. The probability of edge deletion of the grid is set by the size $v_1$ from (7) of the impulse in the initial moment.

The grid reliability for a day ahead of a type "star" is the reliability (Sekine & Imai, 1998) for an edge deletion from the complete graph on six vertices $K_6$. This reliability is

$$p_s = -120 v_1^{15} + 360 v_1^{14} - 270 v_1^{13} - 90 v_1^{12} + 120 v_1^{11} + 20 v_1^9 - 15 v_1^8 - 6 v_1^5 + 1 \qquad (13)$$

The grid reliability for a day ahead of a type "ring" is the reliability (Sekine & Imai, 1998) for an edge deletion from the complete bipartite graph on six vertices $K_{3,3}$, three of which are connected to each of the other three. This reliability is

$$p_t = 79v_1^{12} - 560v_1^{11} + 1668v_1^{10} - 2656v_1^9 + 2331v_1^8 - 960v_1^7 + 96v_1^5 + 21v_1^4 \quad (14)$$

$$-16v_1^3 - 4v_1^2 + 1$$

The grid for a day ahead could be considered as a disordered critical system with a quenched disorder, because the exchanges of EPS mix up in short time. The probability of such critical behaviour of the grid for a day ahead is

$$p_g = 1 - (1/U_s)\exp(-4(\ln U_s - \ln U_p)^2 E_1/(1.261060863\pi)) \quad (15)$$

This probability is obtained from the results (Tsvelik, 2000) for a quenched disorder. The quantity $E_1/(1.261060863\pi)$ has a sense of inverse temperature.

The grid state for a day ahead $s_g$ is

$$s_g = \begin{cases} \text{normal } (n), & \text{if } p_s \neq p_g \text{ and } p_t \neq p_g \\ \text{restorative } (r), & \text{if } p_s = p_g \text{ or } p_t = p_g \\ \text{emergency } (e), & \text{if } p_s = p_g \text{ and } p_t = p_g \end{cases} \quad (16)$$

## 5. Prognostic Watch

The EPS prognostic watch forecasts for a day ahead the threat level of the EPS according to characteristic levels for a day ahead ($T_{6,1}$, $T_{6,2}$, $T_{16}$, $T_{24}$) and according to the droop for a day ahead $k_c$. The grid threat level 'Alarm', which it forecasts according to the market state for a day ahead $s_m$ from (12) and according to the grid state for a day ahead $s_g$ from (16), is

$$\text{Alarm} = \begin{cases} \text{low, if } s_m=n \text{ and } s_g=n \\ \text{guarded, if } s_m=r \text{ and } s_g=n \\ \text{elevated, if } s_m=r \text{ and } (s_g=r \text{ or } s_g=e) \\ \text{high, if } s_m=e \text{ and } s_g=n \\ \text{severe, if } s_m=e \text{ and } (s_g=r \text{ or } s_g=e) \end{cases} \quad (17)$$

**6. Accuracy of the Prognostic Watch**

Let the distances $R_e$, $R_h$, $R_c$, which determine the grid state, are arranged in the following chain $R_2 \leq r \leq R_1$, $R_2 = \min(R_e, R_h, R_c)$, $R_1 = \max(R_e, R_h, R_c)$ and r is the rest of the distances ($R_e$, $R_h$, $R_c$).

The prognostic watch raises false alarm when it expects a hidden threat in the market. The probability for a false alarm is found from the entanglement in a depolarized channel (Majtey, Borras, A. R. Plastino, Casas, & A. Plastino, 2009), representing the market with a hidden threat. Herein, the hidden threat is considered as electromagnetic cloak from Zhau and Hao (2009). The probability for a false threat from the watch is

$$p_f = (2/3)(R_2 / (R_2 - R_1))((r - R_1) / r)^2 \quad (18)$$

Let the probabilities $p_s$, $p_t$, $p_g$, by which is determined the market state, are arranged in the following chain $p_3^* \leq p_2^* \leq p_1^*$. Let the probabilities $p_1$, $p_2$, $p_3$ are set in the following way: $p_1 = p_1^*$, $p_2 = 1 - p_2^*/2$, $p_3 = p_3^*/2$.

The prognostic watch fails to spot a threat, when the grid state can't be determined through quenching. The probability of an omission by the watch is

$$p_m = 1 - 2(p_4/p_3)^{1/2}(((v_m/100)^2(1-v_m/100)^2(p_1-p_2)^2 + p_1p_2)^{1/2} + (p_3p_4)^{1/2}), \quad (19)$$

$$p_4 = (k_c/3.5)^4 p_3$$

This probability is given by the impossibility (Cen, Wu, Yang, & An, 2002) of transforming the mixed state of two qubits to Bell diagonal state.

The mean probability for a false alarm of the Bulgarian EPS prognostic watch is as much as the triple mean relative load forecast error, and the mean probability for the omission by this watch is as much as the mean relative load forecast error. Here the relative error is in relation to the 24 hours peak load and the mean relative load forecast error is 3.5%.

## 7. Conclusion

In this paper is built an EPS prognostic watch, which detects the threat to EPS for a day ahead according to the characteristic times for a day ahead and in relation to the droop for a day ahead. This is done as follows:

(1) The power grid for the day ahead with two values of the expected energy, two expected angular frequencies and two expected times are found. Therefore the EPS Lyapunov exponents are obtained for the day ahead of the expected EPS development;

(2) The state of electricity market for the day ahead and the grid state for a day ahead are found. They are obtained with a prognostic analysis of the potentials in the grid for a day ahead, when the grid is a "star" type and a "ring" type;

(3) EPS level of threat for a day ahead, according to the electricity market state for a day ahead and the grid state for a day ahead, are found.

The accuracy of the prognostic watch is compared to the load forecast error.